\documentstyle[12pt]{article}
\def\bA{{\bar A}}
\def\i{{\frac1{2\pi}\int d^2z\,}} 
\def\ki{{\frac{k}{2\pi}\int d^2z\,}} 
\def\d{\partial}
\def\bd{\bar\partial}

\def\th{\theta}
\def\tL{{\theta_L}}
\def\tR{{\theta_R}}
\def\g{\gamma}

\def\JJ{{\cal{J}}}
\def\bJJ{\bar{\cal{J}}}

\def\de{\delta}

\def\l{\lambda}
\def\s{\sigma}

\def\e{\epsilon}
\def\z{\zeta}
\def\bz{{\bar\zeta}}
\def\bh{{\bar h}}
\def\bJ{{\bar J}}
\def\bK{{\bar K}}
\def\t{\tilde}
\def\implies{\Rightarrow}

\begin{document}

\newcommand{\inv}[1]{{#1}^{-1}} 

\renewcommand{\theequation}{\thesection.\arabic{equation}}
\newcommand{\beq}{\begin{equation}}
\newcommand{\eeq}[1]{\label{#1}\end{equation}}
\newcommand{\ber}{\begin{eqnarray}}
\newcommand{\eer}[1]{\label{#1}\end{eqnarray}}
\begin{center}
        August, 1993
				\hfill    ITP-SB-93-44\\
			        \hfill    RI-152-93\\
                                \hfill    hep-th/9308154\\

\vskip .5in

{\large \bf On Nonabelian Duality}
\vskip .5in

{\bf Amit Giveon} \footnotemark \\

\footnotetext{e-mail address: giveon@vms.huji.ac.il}

\vskip .1in

{\em Racah Institute of Physics, The Hebrew University\\
  Jerusalem, 91904, ISRAEL} \\

\vskip .15in

       and

\vskip .15in

{\bf Martin Ro\v cek} \footnotemark \\

\footnotetext{e-mail address: rocek@insti.physics.sunysb.edu}

\vskip .1in

{\em Institute for Theoretical Physics \\
State University of New York at Stony Brook \\
Stony Brook, NY 11794-3840 USA}\\
\vskip .1in
\end{center}
\vskip .4in
\begin{center} {\bf ABSTRACT } \end{center}
\begin{quotation}\noindent
We show that nonabelian duality is not a symmetry of a conformal field
theory, but rather a symmetry between different
theories.  We expose a nonlocal symmetry of nonabelian dual theories. We
show how, in the case with vanishing isotropy, it can be used to find
the inverse dual transformation. Finally, we consider a number of new
examples.
\end{quotation}
\vfill
\eject
\def\baselinestretch{1.2}
\baselineskip 16 pt
\noindent
\section{Introduction}
\setcounter{equation}{0}

A great deal of insight into the structure of the moduli space of string
backgrounds and conformal field theories (CFT's) has been uncovered by
studies of target space duality (see, for example
\cite{duality,more,GR}). The $O(d,d,{\bf Z})$ symmetry of the CFT's
defined on backgrounds with $d$ toroidal isometries can be realized by
duality transformations of the type studied by T.\ Buscher \cite{TB}.
These are found by gauging a symmetry of the action with nondynamical
gauge fields, {\it i.e.,} without a $F^2$ term, and adding a Lagrange
multiplier that constrains the gauge field to be (locally) pure gauge.
After being careful with the global issues, such dualities can be made
into exact symmetries of the CFT \cite{RV}.

In his thesis, T.\ Buscher \cite{Bth} also considered the possibility of
performing duality transformations for nonabelian
symmetries.\footnote{Earlier examples have been discussed
by Fridling and Jevicki \cite{FJ} and by Fradkin and Tseytlin \cite{FT}.}
More recently, de la Ossa and Quevedo \cite{dOQ} have independently
proposed the same idea, and explored a class of $SO(3)$ examples.
However, they did not consider {\em exact} conformal backgrounds ({\it
e.g.,} WZNW models), and did not analyze the global issues.

Here we consider nonabelian duality for CFT's.  We argue that, except
for ``accidents'', there is no reason to expect nonabelian duality to be
a symmetry of a CFT; at best, it can be a transformation between
different CFT's.  We discover a nonlocal symmetry of dual theories. We
discuss the crucial role of isotropy, and consider several explicit
examples for exact CFT's.

The structure of the paper is as follows: In section {\bf 2}, we begin
with a review of Abelian duality.  In section {\bf 3}, we consider general
aspects of nonabelian duality, and present our argument about the problem
with global issues.  Whereas in the Abelian case, the dual theory has a
natural symmetry suitable for inverting the duality transformation, in
the nonabelian case we find that this symmetry becomes nonlocal and
cannot obviously be used for the inverse. In section
{\bf 4}, we specialize to the case without isotropy, and in section {\bf
5}, we give a few examples.  In section {\bf 6}, we explicitly work out
the $SU(2)$ dual of the $SU(2)$-WZNW model (a case {\em with} isotropy).
We close with some conclusions and a brief discussion.

\section{Review of Abelian Duality}
\setcounter{equation}{0}

In this section, we briefly review Abelian duality
\cite{TB,RV}.  We begin with a target space geometry with a compact
Abelian symmetry.  The duality transformation is performed by gauging this
symmetry and adding a Lagrange multiplier term that constrains the gauge
field to be pure gauge. Integrating out the lagrange multiplier
gives back the original model, and integrating out the gauge field
gives the dual model.

Without loss of generality, we
can choose coordinates on the target space where the symmetry acts by
shifts of a single periodic coordinate $\th\equiv\th+2\pi$, and the
remaining coordinates $x^i$ are left inert. In these coordinates, the
background is independent of $\th$.  The action of the original model
takes the form
\ber
S[\th ,x]=\i \Big( E(x)\d\th\bd\th &+& F^R_j(x)\d\th\bd x^j \ \
+\ \ F^L_i(x)\d x^i\bd\th\nonumber\\
 &+&F_{ij}(x)\d x^i\bd x^j\ -\ \frac14\Phi (x)R^{(2)}
\Big),
\eer{abel}
where $\Phi$ is the dilaton field and $R^{(2)}$ is the world sheet
curvature. We now gauge the symmetry by minimal coupling $\d\th\to\d\th
+A$, $\bd\th\to\bd\th +\bA$, and add a Lagrange multiplier term
\beq
\i ( A\bd\l-\bA\d\l )\ ;
\eeq{Adl}
up to an important total derivative, this is just $\l F$,
$F(A,\bA)\equiv \d\bA-\bd A$. When $\l$ is chosen to have periodicity
$2\pi$, as discussed in \cite{RV}, the winding modes of $\l$ constrain
the holonomy of $A,\bA$ to insure that integrating out $\l$ gives back the
original model.\footnote{In \cite{RV}, though the discussion of
the winding modes of $\l$ is correct, the Lagrange multiplier term was
written as $\l F$, and thus the total derivative term needed to make
the winding modes contribute was accidentally omitted.}

In this form, it is clear that the dual model has a $U(1)$ symmetry that
acts on $\l$ by
\beq
\l\to\l+\e \ ,
\eeq{abelshift}
for $\e$ constant.
If we make a duality transformation with respect to this symmetry, we
immediately see that the dual of the dual is the original model: We
gauge (\ref{abelshift}) and add a second Lagrange multiplier $\t\l$:
\ber
S[x,A,\l ]&\to&
S[x,A,\t A,\l,\t\l ]\nonumber\\
&\equiv& S[x,A,\l ]+\i tr\Big(\t\bA A- \t A\bA +
\t A\bd\t\l-\t\bA\d\t\l\Big).
\eer{abeldudu}
Integrating out $\t A,\t\bA$ gives:
\beq
A=\d\t\l\ ,\qquad \bA=\bd\t\l \ ,
\eeq{puregauge}
and hence we recover the original model.
The inverse transformation is a sign of an underlying {\it group\/} of
duality transformations.  For $d$ commuting $U(1)$ symmetries, one finds an
$O(d,d,{\bf Z})$ group \cite{GR}.

For the case of one symmetry, choosing a gauge $\th =0$, we find the
gauged action
\ber
S[x,A,\l ]=\i\Big( EA\bA&+&F^R_iA\bd x^i+F^L_i\d x^i\bA+F_{ij}\d x^i \bd
x^j \nonumber\\
&{}&\nonumber\\
&-&\frac14\Phi R^{(2)} +(A\bd\l-\bA\d\l )\Big)\ .
\eer{abelgauge}
Integrating out the gauge field $A,\bA$, we find
\beq
A(\l ,x)=(\d\l-\d x^iF^L_i)\inv E\ ,\ \ \
\bA(\l ,x)=-\inv E(\bd\l+F^R_i\bd x^i)\ .
\eeq{abelAbA}
Substituting (\ref{abelAbA}) into the gauged action (\ref{abelgauge}),
we find the dual action:
\ber
\t S[x,\l ]=\i\Big((\d\l-\d x^i F^L_i)\inv E(\bd\l+
F^R_i\bd x^i)\nonumber\\
+ F_{ij}\d x^i\bd x^j -\frac14(\Phi + \ln E)R^{(2)}\Big)\ .
\eer{abeldual}
The shift of the dilaton comes from a Jacobian factor that arises from
integrating over the gauge field $A,\bA$ \cite{TB}.

A general feature of duality is that field equations and Bianchi
identities are rotated into each other \cite{olddual}. Here, they are
simply interchanged.

In the original model (\ref{abel}), the field
equation and Bianchi identity are:
\beq
{\bf Field\ Equation:}\qquad\qquad\qquad\qquad \bd J+\d \bJ = 0\qquad
\qquad\qquad
\eeq{abelfeq}
for
\beq
J=E\d\th+F^L_i\d x^i\ ,\qquad \bJ=E\bd\th+F^R_i\bd x^i \ ,
\eeq{abelJbJ}
and
\beq
{\bf Bianchi\ Identity:}\qquad \d[\inv E(\bJ-F^R_i\bd x^i)]-\bd
[(J-\d x^i F^L_i)\inv E]=0
\eeq{abelBi}
(substituting the definition of $J,\bJ$ (\ref{abelJbJ}) into
(\ref{abelBi}), this becomes just the triviality $\d\bd\th-\bd\d\th=0$).

In the dual model, we have the obvious currents
\beq
\t J=\d\l \ ,\qquad \t{\bJ}=-\bd\l \ .
\eeq{abeltJ}
These obey the dual Bianchi identity:
\beq
{\bf Dual\ Bianchi\ Identity:}\qquad\qquad\qquad\bd\t J +\d\t{\bJ}=0
\ .\qquad\qquad
\eeq{abelduBi}
By construction, the $\l$ field equation is $F(A,\bA )=0$, with
$A,\bA$ given in (\ref{abelAbA}); in terms of the dual currents $\t
J,\t{\bJ}$, this takes the form:
\beq
{\bf Dual\ Field\ Equation:}\qquad \d[\inv E(\t{\bJ}-F^R_i\bd x^i)]-\bd
[(\t J-\d x^i F^L_i)\inv E]=0 \ .
\eeq{abeldufeq}
Thus we see that the field equation (\ref{abelfeq}) becomes the dual
Bianchi identity (\ref{abelduBi}) and the Bianchi identity (\ref{abelBi})
becomes the dual field equation (\ref{abeldufeq}).

\section{General Aspects of Nonabelian Duality}
\setcounter{equation}{0}

As in the Abelian case, we begin by gauging some (non-anomalous) symmetry
of the target space geometry, and then add a Lagrange multiplier
term that constrains the field strength of the gauge field to vanish
\cite{Bth}, \cite{dOQ}:
\beq
S[x^a]\to S[x,A,\l ]\equiv S[x^a,A,\bA]+\i tr (\l F(A,\bA ))\ .
\eeq {gauging}
Here
\beq
F(A,\bA ) \equiv \d \bA - \bd A +[A,\bA ]
\eeq {F}
is the field strength for the world-sheet gauge field $(A,\bA )$ and
$x^a$ are the scalar fields of the original action.  We now consider how
much of our discussion of the Abelian case carries over.

As in the Abelian case, when we integrate out $\l$, we find that
$locally$, $A=h^{-1}dh$ is pure gauge, and hence, modulo global issues,
we recover the original model.

We next consider the question of the holonomy. Here, there are
three immediate and crucial differences. (1) Whereas in the Abelian case
$\l$ is a gauge singlet that can naturally be chosen to be periodic, and
thus acquire the winding modes that act as the Lagrange multipliers for
the holonomies of $A$, in the nonabelian case, $\l$ transforms in the
adjoint representation, and in general the group action is incompatible
with making $\l$ a periodic variable. (2) Even if one found some way to
introduce winding modes for $\l$, they would not multiply the holonomies
of $A$, since in the nonabelian case, these are {\em path-ordered}
exponentials, and hence cannot be written in a local form as terms in the
action. (3) Finally, the $tr\l [A,\bA ]$ term is incompatible with trying
to write the action in a form analogous to (\ref{Adl}).  Consequently,
we conclude that nonabelian duality is $not$
an {\em invariance} of the underlying conformal field theory, but rather
we conjecture that it is a transformation between {\em different}
conformal field theories related by a (infinite order) nonabelian orbifold
construction.\footnote{In the Abelian case, if one neglects to
compactify $\l$, when one integrates it out, one gets $A=d\th$ for
{\em noncompact} $\th$; modding out by the (infinite order) group of
translations of $\th$ by $2\pi$, one recovers the original model.
Equivalently, modding out the original model by the continuous group
$U(1)$ gives rise to continuous twisted sectors that decompactify $\th$.
Similarly, in the nonabelian case, since we cannot compactify $\l$, we
would need to twist the model that we get by integrating out $\l$ (the
``dual CFT") by some infinite order group analogous to the shifts of $\th$
to recover the original model.  Equivalently, we can think of the dual
CFT as the $G$ orbifold of the original CFT, where $G$ is the continuous
group of symmetries that we used for nonabelian duality.}

In the Abelian case, the invariance of the action under constant shifts
of $\l$ is crucial: it guarantees that the dual geometry has a symmetry
that can be used for a further duality transformation, and that the dual
of the dual is the original model (see \ref{abeldudu},\ref{puregauge}).
In the nonabelian case, this symmetry becomes nonlocal, and no longer
serves either function: the dual geometry need not have a local symmetry,
and if one attempts to dualize the nonlocal symmetry, one does not appear
to recover the original model.

To see the nonlocal symmetry, we introduce nonlocal variables to
describe the gauge field:
\ber
h(\z)=h_0 P\ exp({\int^\z Adz})\ &\implies& \ A=h^{-1}\d h\ ,\nonumber\\
&{} &\nonumber\\
\bar h(\bz)=\bar h_0 P\ exp({\int^\bz \bA d\bar z})\ &\implies& \
\bA =\bar h^{-1}\bd \bar h \ ,
\eer {hhbar}
where $P$ denotes path ordering.
Then the field strength $F$ can be rewritten as:
\beq
F(A,\bA )= h^{-1}\d(\inv f \bd f)h=\inv\bh\bd(\d f \inv f)\bh \ ,\qquad
f\equiv \bh\inv h\ .
\eeq {Ff}
This implies that
\beq
tr \l F = tr (h\l\inv h)\d(\inv f\bd f)=
tr (\bh\l\inv\bh )\bd (\d f\inv f) \ ,
\eeq{lFf}
and hence the action (\ref{gauging}) is invariant under the nonlocal
Ka\v c-Moody transformations
\beq
\de \l = \inv h \bar\e (\bar z) h + \inv\bh\e (z)\bh \ .
\eeq {trans}
These are highly nonlocal transformations in terms of the original
fields $A,\bA$ (see \ref{hhbar}), and hence, after $A,\bA$ are
eliminated by their algebraic equations of motion, these are nonlocal in
terms of the coordinates of the dual geometry.

In the Abelian case, we were able to immediately see that the dual of
the dual is the original model by gauging the translation symmetry of
$\l$ and adding a $\t\l F(\t A, \t\bA )$ term (see
\ref{abeldudu}). We can attempt to do the same thing here by gauging the
nonlocal shift symmetry of the action $S[x,A,\l ]$
(\ref{gauging}):
\ber
S[x,A,\l ]&\to& S[x,A,\t A,\l,\t\l ]\nonumber\\
&\equiv& S[x,A,\l ] +\i tr\Big( \t A\inv f \bd f + \t \bA \d f \inv f +
\t\l (\d\t\bA - \bd\t A)\Big). \nonumber\\
\eer{dualdual}
This action is invariant under the nonlocal gauge transformations
\beq
\de\t A = \d \e (z,\bar z) \ ,\qquad \de\t\bA=\bd\e(z,\bar z) \ ,\qquad
\de \l = \inv h \e (z,\bar z) h + \inv\bh\e (z,\bar z)\bh \  .
\eeq {nlgauge}
Integrating out $\t\l$ restricts $\t A,\t\bA$ to be pure gauge (at
least locally), and gives back $S[x,A,\l ]$ after an obvious nonlocal
redefinition of $\l$.  Unfortunately, integrating out $\t A,\t\bA$ does
not yield the original model; in particular, we do not find $F(A,\bA
)=0$ is necessary, and  we get a model from which we cannot eliminate
the various auxiliary degrees of freedom that we have introduced along
the way.

There is of course a simple way to add terms analogous to the Abelian
case (\ref{abeldudu}) to the action $S[x,A,\l ]$ (\ref{gauging}) so that
integrating out $\t A,\t\bA$ gives back the original model:
\ber
S[x,A,\l ]&\to&
S[x,A,\t A,\l,\g ]\nonumber\\
&\equiv& S[x,A,\l
]+\i tr\Big(\t \bA (A-\inv\g\d\g ) -\t A(\bA-\inv\g\bd\g )
\Big)\ . \nonumber\\
\eer{gammadual}
Unfortunately, integrating out $\g$ does not allow us to eliminate $\t
A,\t\bA$; furthermore, after integrating out $A,\bA$ and fixing the
gauge invariance, in general we do not know how to look at the resulting
model and introduce $\t A,\t\bA$ and $\g$.

There is another issue that does not arise in the Abelian case, but is
crucial here: nonabelian gauge groups may have nontrivial $isotropy$
subgroups, that is, they may act on the target space coordinates $x^a$ so
that one cannot completely fix the gauge by conditions on $x^a$.  (For
example, this is the case for the vector action of a group $G$ on the
$G$-WZNW model; for another class of examples, see \cite{dOQ}.)  In this
case, in the dual model, some of the dual coordinates $\l$ must be gauge
fixed, and a clear separation into original and dual coordinates is
impossible.

\section{The case without isotropy}
\setcounter{equation}{0}

When the isotropy vanishes, {\it i.e.\/}, when the gauge symmetry can be
fixed completely by conditions on the coordinates of the original model,
some more aspects of nonabelian duality resemble the Abelian case.
This is the generic situation for WZNW models: For a group $G$
dualized with respect to a proper subgroup $H$, {\it e.g.} $G=SU(N)$,
$H=SU(N-m)$, $m\neq 0$, the isotropy vanishes at generic points of $G$; on
certain singular subspaces, it becomes nonvanishing; at these points,
the dual space has singularities, which, however, are not peculiar from
the viewpoint of the underlying CFT.

\subsection{The original model.}

In general, we can consider a target space
with coordinates $g$ that transform as $g\to\inv u g$ for $u$ in some
group $G$, and further coordinates $x^i$ that are inert.\footnote{Any
geometry with vanishing isotropy can be parametrized  by coordinates $g$
and $x^i$ because when the isotropy vanishes, we can fix the gauge
completely by algebraic conditions on the coordinates of the target
space.  We call the remaining coordinates, whatever they are, $x^i$.  We
then transform to a general gauge with a parameter $g$ which becomes our
coordinate $g$. By the group property, subsequent (passive) gauge
transformations then transform $g$ correctly and leave $x^i$ inert. For
the example $G/H$ mentioned above, we would write an element $g_G\in G$
in terms of $g\equiv g_H\in H$ and $x$ as $g_G=g_Hx\inv g_H$.} A general
non-anomalous action can be written in the form
\ber
S[g,x]=\i \Big( E_{ab}(x)(\inv g\d g)^a(\inv g\bd g)^b &+&
F^R_{aj}(x)(\inv g\d g)^a\bd x^j \nonumber\\
+\ \ F^L_{ib}(x)\d x^i(\inv g\bd g)^b &+&F_{ij}(x)\d x^i\bd
x^j \nonumber\\
&-&\frac14\Phi (x)R^{(2)}\Big)\ ,
\eer {noiso}
where
\beq
(\inv g\d g)^a \equiv tr (T^a \inv g\d g) \ \ \Leftrightarrow \ \ \inv
g\d g = (\inv g\d g)^a T_a\ ,\ \ etc.,
\eeq{gdga}
and the generators $T_a$, $a=1,...,dim(G)$, obey
\beq
tr(T_a T_b)= \eta_{ab}\
\eeq{trTT}
for some invariant metric $\eta$ that can be used to raise and lower
indices.  Under an infinitesimal variation $\de g=\e^aT_ag$, we find the
field equation
\beq
\d \bJ +\bd J =0\ ,
\eeq{feq}
where the currents $J,\bJ$ are
\ber
J=g\JJ\inv g\ ,&\ \ & \bJ=g\bJJ\inv g\ ,\nonumber\\
\JJ^a=(\inv g\d g)^bE_b{}^a+\d x^iF_i^{La} \ ,&\ \ &
\bJJ^a=E^a{}_b(\inv g\bd g)^b+F_i^{Ra}\bd x^i\ .
\eer{JbJ}
The currents $J,\bJ$ obey Bianchi identities which follow from
\beq
F(\inv g\d g,\inv g\bd g)=0\ ,
\eeq{Bi}
where $F$ is the field strength (\ref{F}).
In terms of $J,\bJ$, this involves explicit factors of $g$, and hence
the form is explicitly gauge dependent; this can be cured by rewriting the
identities in terms of the conjugated currents $\JJ,\bJJ$. They become
just
\vskip .1in
${\bf Bianchi\ Identity:}\ \qquad\qquad  F(V(g,x),\bar V(g,x))=0$,
\beq
V^a(g,x)=(\JJ^b-\d x^iF_i^{Lb})E^{-1}_b{}^a \ , \qquad
{\bar V}^a(g,x)=E^{-1a}{}_b(\bJJ^b-F^{Rb}_i\bd x^i) \ .
\eeq{JJBia}
Now the field equations (\ref{feq}) become:
\beq
{\bf Field\ Equation:}\ \qquad\qquad \d\bJJ+\bd\JJ+[V,\bJJ ]+[\bar V,\JJ
]=0\ . \qquad
\eeq{JJfeq}

\subsection{The dual model.}

To find the dual, we gauge the action $S[g,x]$ (\ref{noiso}) by minimal
coupling:
\beq
\inv g\d g \to \inv g(\d +A)g\ ,\ \ \
\inv g\bd g \to \inv g(\bd +\bA)g\ .
\eeq{mincoup}
Here $A$ transforms as $A\to \inv u(\d +A)u$, {\it etc.} Adding the $\l
F$ term, and choosing the gauge $g=1$, we get the action
\ber
S[x,A,\l ] = \i \Big( E_{ab}A^a\bA^b&+&F^R_{ai}A^a\bd x^i+F^L_{ia}\d
x^i\bA^a +F_{ij}\d x^i\bd x^j \nonumber\\
-\ \ \frac14\Phi R^{(2)}&+&\l_a(\d\bA^a-\bd A^a +f^a_{bc} A^b\bA^c)
\Big) \ .
\eer{SxA}
Integrating out the gauge fields $A,\bA$, we get
\ber
A^a(\l ,x)&=&(\d\l_b-\d x^iF^{L}_{ib})[(E+\l_cf^c)^{-1}]^{ba}\ ,\nonumber\\
{}&{}&{}\nonumber\\
\bA^a(\l ,x)&=&-[(E+\l_cf^c)^{-1}]^{ab}(\bd\l_b+F^{R}_{bi}\bd x^i)\ ,
\eer{AbA}
where the matrices $f^c$ have the structure constants as components
$(f^c)_{ab}=f^c_{ab}$. We find the dual theory
\ber
\t S[x,\l ]=
\i \Big( (\d\l_b-\d x^iF^{L}_{ib})
[(E+\l_cf^c)^{-1}]^{ba}(\bd\l_a+F^{R}_{aj}\bd x^j)\nonumber\\
+\ \ F_{ij}\d x^i\bd x^j\ -\ \frac14[\Phi+\ln (\det
[E+\l_cf^c])]R^{(2)}\Big)\ ,
\eer{Sdual}
where the transformation of the dilaton comes from the
Jacobian as in (\ref{abeldual}).

In the Abelian case, duality interchanges the field equations and the
Bianchi identities (\ref{abelBi}); it is instructive to see what happens
here.

The dual field equations (from the variation of $\l$) are precisely
\beq
{\bf Dual\ Field\ Equation:}\ \qquad\qquad
F(A(\l ,x),\bA (\l ,x))=0\qquad
\eeq{dfeq}
for $A,\bA $ given by (\ref{AbA}); this has
exactly the same form as the Bianchi identity (\ref{JJBia}) in the
original model, and suggests that we identify ``currents" $\t\JJ
,\t{\bJJ}$:
\beq
\t\JJ^a(\l ,x)=A^bE_b^{\ a}+\d x^iF^{La}_i \ , \ \ \
\t{\bJJ}^a(\l ,x)=E^a_{\ b}\bA^b+F^{Ra}_i\bd x^i \ ,
\eeq{tJJ}
with $A,\bA$ given by (\ref{AbA}).  The dual ``Bianchi" identity is the
triviality $\d\bd\l=\bd\d\l$; however, when this is written in terms of
$\t\JJ,\t{\bJJ}$, it becomes nontrivial.
Using (\ref{AbA},\ref{tJJ}), we write $\d\l ,\bd\l$ as
\beq
\d\l=\t\JJ+[\l ,A]
\ ,\ \ \
\bd\l=-\t{\bJJ}+[\l ,\bA ]\ ,
\eeq{dlbdl}
and then find
\beq
{\bf Dual\ Bianchi\ Identity:}\ \ \
\d\t{\bJJ}+\bd\t\JJ+[A,\t{\bJJ}]+[\bA
,\t\JJ ]=[\l ,F(A,\bA )]\ .
\eeq{dBia}
The right hand side is proportional to the dual field equation
(\ref{dfeq}); the remaining term has the form of the field equation
(\ref{JJfeq}) in the original model.

\subsection{Some comments on the duality group.}

In the Abelian case, when there is more than one isometry, one finds a
{\it duality group} \cite{duality,GR}.  Some of the generators of this
group act by adding a total derivative to the Lagrangian; this shifts
the antisymmetric tensor (Kalb-Ramond field).  Here we will find certain
shifts of a similar type that lead to dual backgrounds that differ
merely by coordinate transformations (shifts of $\l$).  Indeed, the
dual action $\t S[x,\l ]$ (\ref{Sdual}) has the following symmetry:
\beq
\l\to\l -\xi (x)\ ,
\eeq{lshift}
\beq
E_{ab}\to E_{ab}+\xi_cf^c_{ab}\ ,
\ \ \ F^{R}_{ai}\to F^{R}_{ai}+\frac{\d}{\d x^i}\xi_a \ ,
\ \ \ F^{L}_{ia}\to F^{L}_{ia}-\frac{\d}{\d x^i}\xi_a \ .
\eeq{shift}
In particular, for $\xi$ constant, this is just a constant shift of
$E_{ab}$. In general, this transformation of the background fields
$E,F^R,F^L$ shifts the original action (\ref{noiso}) by the total
derivative term
\beq
\i\Big(\bd[\xi_a(\inv g\d g)^a]-\d[\xi_a(\inv g\bd g)^a] \Big)
\eeq{totder}
(we use the Bianchi identity $F(\inv g\d g,\inv g\bd g)=0$).

One of the outstanding problems in nonabelian
duality is the nature of the group of duality transformations.  As
discussed in section {\bf 3}, since, in general, nonabelian duality
transformations take backgrounds with symmetries into background without
corresponding symmetries, we do not even know how to perform the inverse
transformation.  In the case without isotropy, however, it seems that
there is a way of finding the inverse using the nonlocal
transformations (\ref{trans}).  One might hope that the transformation
(\ref{trans}) uniquely defines the coordinates $\l$ precisely up to shifts
(\ref{lshift}); as discussed above, these shifts do not change
the original (``anti-dual") background.

When the $g$'s are elements of, {\it e.g.,} $G\times G\times \dots
\times G=G^d$, then we can look for a duality group analogous to $O(d,d)$
when $G$ is $U(1)$; this is currently under investigation.

\section{An example without isotropy}
\setcounter{equation}{0}

The analysis of the previous section did not address the issue of
world sheet conformal invariance. A simple example of nonabelian duality
without isotropy for conformal field theories arises for the WZNW model on
$G\times G$ at level $k$ dualized with respect to a particular
nonanomalous $G$ subgroup. Explicitly, we start with  \beq
S_{G_k\times G_k}[g_1,g_2]=S_{G_k}[g_1]+S_{G_k}[g_2] \ ,
\eeq{SGG}
where $S_{G_k}$ is the WZNW action for $G$ at level $k$:
\beq
S_{G_k}\equiv\frac{k}{2\pi}\Big[\frac12\int d^2z\, tr(\inv g \d g \inv g
\bd g) +\Gamma\Big]\ ,
\eeq{SGk}
where $\Gamma$ is the Wess-Zumino term. We gauge the subgroup \cite{EW}
\beq
g_1\to \inv u g_1 \ ,\qquad g_2\to g_2 u \ .
\eeq{GGsym}
The gauged action is
\ber
S_{gauge}&=&S_G[fg_1]+S_G[g_2\bar f]-S_G[f\bar f]\nonumber\\
&{}&\nonumber\\
&=&S_G[g_1]+S_G[g_2]+\ki tr\Big(A\bd g_1 \inv
{g_1}-\inv{g_2}\d g_2\bA+ A\bA \Big)\ ,\nonumber\\
\eer{GGgauge}
where $A\equiv\inv f \d f$ and $\bA\equiv -\bd{\bar f}\inv{\bar f}$,
and we have used the Polyakov-Wiegmann formula \cite{PW}
\beq
S_{G_k}[gh]=S_{G_k}[g]+S_{G_k}[h]+\ki tr\Big(\inv g\d g \bd h \inv
h\Big)\ .
\eeq{poly}
We then add the usual $\l F(A,\bA)$ term.

With the coordinates $g_1,g_2$, the action $S_{G\times G}$ (\ref{SGG})
does not manifestly have the structure of the models discussed in the
previous section; to find this structure, we define new variables $g,x$,
with $x$ inert under (\ref{GGsym}), as follows:
\beq
g\equiv g_1\ ,\qquad x\equiv g_2g_1 \ .
\eeq{gxdef}
Then using Polyakov-Wiegmann formula (\ref{poly}),
the action $S_{G\times G}$ takes the form
\beq
S_{G\times G}[g,x]=S_G[x]+\ki tr \Big(\inv g\d g \inv g\bd g - \inv x\d
x\inv g\bd g\Big) \ .
\eeq{SGGx}
This action has the form (\ref{noiso}) with $E_{ab}=k\eta_{ab}$,
$F^L_{ia}=-ke_{ia}$ where $e$ is the usual left-invariant 1-form of the
group, $F^R_{ai}=0$, and $F_{ij}(x)$ is the usual background for a
$G$-WZNW model. In this form, the gauging (\ref{GGgauge}) is given simply
by the minimal coupling prescription (\ref{mincoup}), and, after choosing
the gauge $g=1$, takes the form  (\ref{SxA}).

For $G=SU(2)$, the resulting dual space is a six-dimensional space with
three noncompact coordinates $\l$ fibering over $SU(2)$ (coordinatized
by $x$). Such a background could arise in compactifying strings.
It would be interesting to investigate its
geometry.

\section{The $G$-dual of the $G$-WZNW model: An example with isotropy}
\setcounter{equation}{0}

\subsection{General discussion}

The simplest examples of $G$-WZNW models with isotropy are the duals
with respect to the full group $G$. We begin by gauging $S_G$ in the
standard way \cite{GWZNW}:
\ber
S_G[g]&\to& S_{G/G}[g,f,\bar f]=S_G[fg\bar f]-S_G[f\bar f]\nonumber\\
\nonumber\\
&=&S_G[g]+\ki tr \Big(A\bd g\inv g - \inv g \d g \bA +A\bA - Ag\bA\inv g
\Big) \ ,\nonumber\\
\eer{isogauge}
where $A\equiv\inv f \d f$ and $\bA\equiv -\bd{\bar f}\inv{\bar f}$,
and we have used the Polyakov-Wiegmann formula (\ref{poly}).  This
action is invariant under
\beq
g\to\inv u g u\ ,\ \ \ f\to fu\ ,\ \ \ \bar f\to \inv u \bar f\ .
\eeq{isotrans}
These transformations do not allow us to reach the gauge $g=1$, and
hence the isotropy doesn't vanish.  The best we can do is, {\it e.g.,}
to choose $g$ to lie in the Cartan torus.\footnote{This manifests itself
in that the $G/G$ CFT is a {\em nontrivial} topological field theory.}
To find the nonabelian $G$-dual of $G$, we add the standard $\l F$ term;
further gauge choices to completely fix the gauge invariance can be made
on the Lagrange multipliers $\l$ themselves \cite{dOQ}.

Before gauge fixing, the total action becomes
\beq
S=S_G[g]+\i\Big(A^a\bK_a-K_a\bA^a+A^aM_{ab}\bA^a\Big) \ ,
\eeq{isoStot}
where
\beq
K=k\inv g\d g+\d\l\ ,\ \ \ \bK=k\bd g\inv g + \bd\l\ ,\ \ \
M_{ab}=k[\eta_{ab}-tr(\inv g T_a g T_b)]+\l_cf^c_{ab}\ ,
\eeq{isoK}
and indices are raised and lowered by the Killing metric
$\eta_{ab}=tr(T_aT_b)$. Integrating out the gauge fields, we find
\beq
\t S[g,\l ]=S_G[g]+\i (K_a[\inv M]^{ab}\bK_b -\frac14\ln(\det
[M])R^{(2)})\ .
\eeq{isoSdual}
This action is still gauge invariant, and may be gauge fixed by
eliminating certain components of $g$ and $\l$.

\subsection{The $SU(2)$ example}

For $SU(2)$, we parametrize the group element $g$ as:
\beq
g=e^{\frac{i}2\tL\s_3}e^{\frac{i}2\phi\s_2}e^{\frac{i}2\tR\s_3}
=\left(\matrix{e^{\frac{i}2(\tL+\tR)}cos\frac\phi2\
{}&{}\ e^{\frac{i}2(\tL-\tR)} sin\frac\phi2\cr{}&{}\cr
-e^{-\frac{i}2(\tL-\tR)}sin\frac\phi2\ {}&{}\
e^{-\frac{i}2(\tL+\tR)}cos\frac\phi2}\right) \ , \eeq{isog}
where $\s_a$ are the Pauli matrices and
\beq
\frac12(\tL\pm\tR )\in[0,2\pi
),\phi\in[0,\pi)
\eeq{isoeu}
are Euler angles. We normalize the generators as
$T_a\equiv\frac{i}2\s_a$. Under an infinitesimal transformation $\de
g=[g,\e ]$, $\e\equiv\e^aT_a$, the Euler angles transform as:
\ber
\de\phi&=&-\e_1(sin\tL+sin\tR)+\e_2(cos\tL-cos\tR)\nonumber\\
\nonumber\\
\de\tR&=&\e_3+\frac{\e_1}{sin\phi}(cos\tL-cos\phi
cos\tR)+\frac{\e_2}{sin\phi}(sin\tL+cos\phi sin\tR )\nonumber\\
\nonumber\\
\de\tL&=&-\e_3+\frac{\e_1}{sin\phi}(cos\tR-cos\phi
cos\tL)-\frac{\e_2}{sin\phi}(sin\tR+cos\phi sin\tL )\ .\nonumber\\
\eer{isoeutrans}
Note that, {\it e.g.,} $\e_1$ and $\e_3$ can be used to go to a gauge
$\tR=\tL=0$, but that in that gauge, $\e_2$ does not act on any of the
Euler angles. This is an explicit demonstration of the nontrivial
isotropy of this gauging.

We now turn to the Lagrange multiplier $\l\equiv\l^aT_a$; it transforms
as $\de\l =[\l ,\e ]$, or, explicitly
\beq
\de\l_1=\l_2\e_3-\l_3\e_2\ ,\ \ \de\l_2=\l_3\e_1-\l_1\e_3\ ,\ \
\de\l_3=\l_1\e_2-\l_2\e_1\ .
\eeq{isoltrans}
We can fix the $\e_2$ transformation by choosing, {\it e.g.,} a gauge
$\l_3=0$.

Now, evaluating the dual action (\ref{isoSdual}) for $SU(2)$ with the
particular gauge choice $\tR=\tL=\l_3=0$, we find a $\s$-model with no
antisymmetric tensor and with a metric
\beq
ds^2=\frac{k}4\Big((d\phi)^2+\frac1{\l_1^2sin^2\frac\phi2}
[4sin^4\frac\phi2(d\l_2)^2+
(\frac1k\l_1d\l_1+[\frac1k\l_2+\phi-sin\phi]d\l_2)^2]\Big)\ .
\eeq{isometric}
We also find a dilaton
\beq
\Phi=\ln(\l_1^2sin^2\frac\phi2) \ .
\eeq{isodil}
We have checked {\em explicitly} that this background solves the one-loop
$\beta$-function equations (for a review, see \cite{GSW}):
\beq
R_{ab}=\nabla_a\nabla_b\Phi \ .
\eeq{oneloop}
The singularities at $\l_1=0$ and at $\phi=0$ clearly arise because our
gauge choice breaks down at these points, and hence are an artifact of
the $\s$-model description of the CFT.

\section{Comments and Discussion}
\setcounter{equation}{0}

In this paper, we have discussed aspects of nonabelian duality.  We
found that it is {\em not} an exact symmetry of the conformal field
theory, and conjectured that it is a map between {\em different} conformal
field theories. We also found that, though in general nonabelian duality
does not preserve the isometries of the original target space, the dual
theory has a nonlocal symmetry that somehow encodes the information
about its more symmetric progenitor.  We emphasized the key role of
isotropy, and found that most WZNW models have vanishing isotropy at
generic points.  For such models, we sketched how the nonlocal
symmetry could be used to find the inverse nonabelian dual
transformation.  We were not able to solve the outstanding problem of
finding the group structure of nonabelian duality transformations.
We also worked out several examples in some detail, both with and
without isotropy.

\vskip .3in \noindent
{\bf Note Added} \vskip .2in \noindent
We thank the referee for suggesting that we expand the discussion of
whether the dual of a conformal field theory is conformal.  We argue
that this is the case as long as the gauging of the isometry does not
introduce any anomalies that cannot be cancelled by modified
transformations of the Lagrange-multiplier fields $\l$. Our argument runs
as follows: whether a theory is conformally invariant has to do with
the short distance behavior of the theory.  In particular, if it is
conformal on a topologically trivial world sheet (modulo possible
global anomalies), it is conformal; however, on a topologically trivial
world sheet, the nonabelian dual of a theory is the same as the original
theory; non-trivial flat connections arise only on world sheets that are
not simply connected.  Hence, if the gauging procedure itself does not
break conformal invariance, the nonabelian dual of a conformal field
theory must be conformal. In this work we described examples that fall into
this category, namely, (non-anomalous) gauged WZNW models.

After the completion of our original manuscript, \cite{GRV} appeared.
It includes an example of a CFT whose nonabelian dual is {\em
not} conformal; however, the group gauged in that example has structure
constants that are {\em not} traceless ($f^a_{ab}\ne 0$); this gives
rise to an anomaly (not the standard one), and hence the argument given
above does not apply.

\vskip .3in \noindent
{\bf Acknowledgements} \vskip .2in \noindent
We would like to thank Warren Siegel for discussions, and the Institute
for Theoretical Physics at Santa Barbara, where part of this work was
done, for its warm hospitality. AG would like to thank Elias Kiritsis
and Gabriele Veneziano for discussions, and the Institute for
Theoretical Physics at Stony Brook, where part of this work was done, for
its warm hospitality. This work was supported in part by NSF grant No.
PHY 9211367.

\end{document}